\documentstyle[12pt]{article}
\newcommand{\beq}{\begin{equation}}
\newcommand{\eeq}{\end{equation}}

\begin{document}
\begin{center}

{\bf Calculations of single particle spectra in density functional
theory}\bigskip\\

M. Ya. Amusia$^{a,b}$, V. R. Shaginyan$^{c,}$ \footnote{E--mail:
vrshag@thd.pnpi.spb.ru}\bigskip

{\it $^{a}$The Racah Institute of Physics, Hebrew University,
Jerusalem 91904, Israel\\[0pt] $^{b}$A.F. Ioffe Physical-Technical
Institute, St. Petersburg 194021, Russia \\[0pt] $^{c}$Petersburg
Nuclear Physics Institute , Gatchina 188350, Russia}\\[0pt]
\end{center}

\begin{abstract}
In the context of the density functional theory we consider the
single particle excitation spectra of electron systems. As a
result, we have related the single particle excitations with the
eigenvalues of the corresponding Kohn-Sham equations. In a case
when the exchange correlation functional is approximated by the
exchange functional the coupling equations are very simple, while
the single particle spectra do not coincide neither with the
eigenvalues of the Kohn-Sham equations nor with the ones of the
Hartree-Fock equations.
\end{abstract}

\noindent {\it PACS:} 31.15.Ew; 31.50.+w
\noindent {\it Keywords:} Density functional theory; Effective
interaction; Excitation spectra

\vspace{0.3cm}

The density functional theory (DFT), that originated from the
pioneer work of Hohenberg and Kohn \cite{wks}, has been extremely
effective in describing the ground state of finite many- electron
systems. Such a success gave birth to many papers concerned with
generalizations of DFT, which would permit to describe the
excitation spectra also ( see e.g. [2 - 5]). The
generalizations, on the theoretical grounds, originated mainly
from the Runge-Gross theorem, which helped to transform DFT into
the time-dependent density functional theory TDDFT \cite{rg}.
Both, DFT and TDDFT, are based on the one-to-one correspondence
between particle densities of the considered systems and external
potentials acting upon these particles. For the sake of simplicity
and definiteness, while dealing with TDDFT, we assume that the
time dependent part of the considered electronic system's density
$\rho({\bf r},t)$, which is created under the action of an
external time-dependent field $\lambda v_{ext}({\bf r},t)$, is
developing from the system's ground state. Unfortunately, the
one-to-one correspondence establishes only the existence of the
functionals in principle, leaving aside a very important question
on how one can construct them in reality. This is why the
successes of DFT and TDDFT strongly depend upon the availability
of good approximations for the functionals. This shortcoming was
resolved to large extent in \cite{ksk,s1,as} where an exact
equations connecting the action functional, effective interaction
and linear response function were derived. But the linear response
function, containing information of the particle-hole and
collective excitations, does not directly present information
about the single particle spectrum.

The main goal of this Letter is to derive  equations describing
single particle excitations of multi-electron systems, using as a
basis the exact functional equations. As a result, by constructing
more and more accurate functionals, one will among other things
treat more and more accurately the single particle excitations.

Let us briefly outline the equations for exchange-correlation
functional $E_{xc}[\rho]$ of the ground state energy and
exchange-correlation functional $A_{xc}[\rho]$ of the action
$A[\rho]$ in the case when the system in question is not perturbed
by an external field. In that case an equality holds \beq
E_{xc}[\rho]=A_{xc}[\rho]|_{\rho(r,\omega=0)},\eeq since $A_{xc}$
is also defined in the static densities domain. The
exchange-correlation functional $E_{xc}[\rho]$ is defined by
the total energy functional $E[\rho]$ (see e.g. \cite{gdp}) as
\beq E[\rho]\equiv T_0[\rho] +\frac{1}{2}\int \frac{\rho({\bf
r}_1)\rho({\bf r}_2)} {|{\bf r}_1-{\bf r}_2|}d{\bf r}_1d{\bf
r}_2+E_{xc}[\rho],\eeq where $T_0[\rho]$ is the functional of
non-interacting Kohn-Sham particles. The atomic system of units
$e=m=\hbar=1$ is used in this paper. The exchange-correlation
functional may be obtained from:
\beq E_{xc}=
-\frac{1}{2}\int_0^1 dg'\int \left[ \chi({\bf r}_1,{\bf
r}_2,iw,g')+ 2\pi\rho({\bf r}_1)\delta(w)\delta({\bf r}_1-{\bf
r}_2)\right] \frac{1}{|{\bf r}_1-{\bf r}_2|}
\frac{dw}{2\pi}d{\bf r}_1d{\bf r}_2. \eeq
Equation (3) presents the well-known expression for the
exchange-correlation energy of a system ( see e.g.
\cite{ksk,as,pp}), expressed via linear response function
$\chi({\bf r}_1,{\bf r}_2,iw,g')$, with $g'$ being the coupling
constant. In order to consider (3) as describing $E_{xc}$, the
only thing we need, is the ability to calculate the functional
derivatives of $E_{xc}$ with respect to the density. According to
(3), it means an ability to calculate the functional derivatives of
the linear response function $\chi$ with respect to the density
$\rho({\bf r},\omega)$ which was developed in \cite{ksk,as}. The
linear response function is given by the integral equation, \beq
\chi({\bf r}_1,{\bf r}_2,\omega,g)= \chi_0 ({\bf r}_1,{\bf
r}_2,\omega)+ \int \chi_0 ({\bf r}_1,{\bf r}'_1,\omega) R({\bf
r}\,'_1,{\bf r}\,'_2,\omega,g) \chi({\bf r}'_2,{\bf r}_2,\omega,g)
d{\bf r}'_1 d{\bf r}'_2, \eeq with $\chi_0$ being the linear
response function of non-interacting Kohn-Sham particles, moving
in the single particle time-independent field \cite{ksk,as}. It is
evident that the linear response function $\chi(g=1)$ tends to the
linear response function of the system in question as $g$ goes to
1. The exact functional equation for $R({\bf r}_1,{\bf
r}_2,\omega,g)$ \cite{ksk,as} looks as follows, \beq R({\bf
r}_1,{\bf r}_2,\omega,g) =\frac{g}{|{\bf r}_1-{\bf r}_2|}\eeq
$$-\frac{1}{2} \frac{\delta^2}{\delta\rho({\bf r}_1,\omega)
\delta\rho({\bf r}_2,-\omega)} \int\int_0^g\chi({\bf r}_1',{\bf
r}_2',iw,g') \frac{1}{|{\bf r}_1'-{\bf r}_2'|} d{\bf r}_1\,'d{\bf
r}_2\,'\frac{dw}{2\pi}\,dg'.$$ Here $R({\bf r}_1,{\bf
r}_2,\omega,g)$ is the effective interaction depending on the
coupling constant $g$ of the Coulomb interaction. The coupling
constant $g$ in eq. (5) is varied in the range $(0-1)$. The
single particle potential $v_{xc}$, being time-independent, is
determined by the following relation \cite{ksk,as}, \beq
v_{xc}({\bf r}) =\frac{\delta}{\delta\rho({\bf r})}E_{xc}. \eeq
Here the functional derivative is calculated at $\rho=\rho_0$ with
$\rho_0$ being the equilibrium density. By substituting (3) into
(6), it can be shown that the single particle potential $v_{xc}$
has the proper asymptotic behavior \cite{as,ab}, \beq
v_{xc}(r\to\infty)\to v_{x}(r\to\infty)\to-\frac{1}{r}.\eeq The
potential $v_{xc}$ determines the energies $\varepsilon_i$ and
wave functions $\phi_i$ of fictitious Kohn-Sham particles, \beq
\left(-\frac{\nabla^2}{2}+V_H({\bf r})+V_{ext}({\bf r})
+v_{xc}({\bf r}) \right)\phi_i({\bf r})
=\varepsilon_i\phi_i({\bf r}), \eeq 
that in turn form the linear response function $\chi_0$, \beq
\chi_0({\bf r}_1,{\bf r}_2,\omega)=\sum_{i,k}
n_i(1-n_k)\phi^*_i({\bf r}_1)\phi_i({\bf r}_2) \phi^*_k({\bf
r}_2)\phi_k({\bf r}_1) \left[\frac{1}{\omega-\omega_{ik}+i\eta}
-\frac{1}{\omega+\omega_{ik}-i\eta}\right], \eeq and the real
density of the system $\rho$, \beq \rho({\bf r})=\sum_i
n_i|\phi_i({\bf r})|^2. \eeq Here $n_i$ are the occupation
numbers, $V_{ext}$ contains all external single particle
potentials of the system, say the Coulomb potentials of the
nuclei. Then, $V_H$ is the Hartree potential, \beq V_H({\bf
r})=\int\frac{\rho({\bf r}_1)} {|{\bf r}-{\bf r}_1|} d{\bf
r}_1,\eeq and $\omega_{ik}$ is the one-particle excitation energy
$\omega_{ik}=\varepsilon_k-\varepsilon_i$, $\eta$ being an
infinitesimally small positive number.
A few remarks regarding the well-known paradox
\cite{gdp,bg} are in order here. The linear response function
$\chi$ and its inverse $\chi^{-1}$ are the noncausal function,
with the inverse being given by \cite{ksk,as}, \beq \chi^{-1}({\bf
r}_1,{\bf r}_2,\omega) =\frac{\delta^2 A[\rho]}{\delta\rho({\bf
r}_1,\omega) \delta\rho({\bf r}_2,-\omega)}.\eeq These functions
can be used on equal footing with the causal ones without any
contradictions related to the interplay between the causality and
symmetry properties of the linear response functions
\cite{ksk,asl,hb}. While the inverse, given by eq. (12), can be
used to construct both causal and noncausal linear response
functions. The described above equations (2-5) solve the problem
of calculating $E_{xc}$, the ground state energy and the
particle-hole and collective excitation spectra of a system
without resorting to approximations for $E_{xc}$ based on
additional and foreign to the considered problem calculations such
as Monte Carlo simulations, or something of this kind. We note,
that using these approximations, one faces difficulties in
constructing the effective interaction of finite radius and the
linear response functions \cite{gdp,bg}. On the base of the
suggested approach one can solve these problems. For instance, in
the case of a homogeneous electron liquid it is possible to
determine analytically an efficient approximate expression
$R_{RPAE}$ for the effective interaction $R$, which essentially
improves the well-known Random Phase Approximation \cite{s1,as} by
taking into account the exchange of electrons properly, thus
forming the Random Phase Approximation with Exchange. The
corresponding expression for $R_{RPAE}$ is as follows \beq
R_{RPAE}(q,g,\rho) =\frac{4\pi g}{q^2}+R_E(q,g,\rho), \eeq where
\beq R_E(q,g,\rho)=-\frac{g\pi}{p_F^2}\left[\frac{q^2}{12p_F^2}
\ln\left|1-\frac{4p_F^2}{q^2}\right| -\frac{2p_F}{3q}\ln
\left|\frac{2p_F-q}{2p_F+q}\right|+\frac{1}{3}\right]. \eeq Here
the electron density $\rho$ is connected to the Fermi momentum by
the ordinary relation $\rho=p^3_F/3\pi^2$. Having at hand the
effective interaction $R_{RPAE}(q,g,\rho)$, one can calculate the
correlation energy $\varepsilon_c$ per electron of the electron
gas with the density $r_s$. The dimensionless parameter
$r_s=r_0/a_B$ is usually introduced to characterize the density,
with $r_0$ being the average distance between electrons, and $a_B$
is the Bohr radius. The density is high, when $r_{s}\ll1$.\\ Table
1. \hfill\parbox[t]{14cm}{Correlation energy per electron in eV of
an electron gas of the density $r_s$. The Monte Carlo results
\cite{mc} $\varepsilon^M_c$ are compared with the RPA calculations
and with the results of \cite{s1}. $\varepsilon_{RPA}$ denotes the
results of RPA calculations, and $\varepsilon_{RPAE}$ denotes the
results of the calculations when the effective interaction $R$ was
approximated by $R_{RPAE}$ \cite{s1}.}
\begin{center}\bigskip
\begin{tabular}{|r|l|l|l|}  \hline\hline
$r_s$ & $\varepsilon^M_c$ & $\varepsilon_{RPA}$ &
$\varepsilon_{RPAE}$
\\ \hline\hline
1 & -1.62 & -2.14 & -1.62 \\ \hline 3 & -1.01 & -1.44 & -1.02 \\
\hline 5 & -0.77 & -1.16 & -0.80 \\ \hline 10 & -0.51 & -0.84 &
-0.56 \\ \hline 20 & -0.31 & -0.58 & -0.38 \\ \hline 50 & -0.16 &
-0.35 & -0.22\\  \hline\hline
\end{tabular}
\end{center}\bigskip
As can be seen from Table 1, the effective interaction
$R_{RPAE}(q,\rho)$ permits to describe the electron gas
correlation energy $\varepsilon_c$ in an extremely broad interval
of density variation. Note, that even at $r_s=10$ the mistake is
no more than 10\% as compared to Monte Carlo calculations, while
the result becomes almost exact at $r_s=1$ and is absolutely exact
when $r_s\to 0$ \cite{s1}.

Now let us calculate the single particle energies $\epsilon_i$,
that, generally speaking, do not coincide with the eigenvalues
$\varepsilon_i$ of eq. (8). Note that these eigenvalues do not
have the physical meaning of a particle energy (see e.g.
\cite{wks}). To calculate the single particle energies one can use
the Landau equation \cite{ll} \beq \frac{\delta E}{\delta
n_i}=\epsilon_i.\eeq Eq. (15) can be used since, as it follows
from eqs. (4,9,10), the linear response functions $\chi$ and
$\chi_0$, and the density $\rho$ depend upon the occupation
numbers. Thus, one can consider the ground state energy as a
functional of the density and the occupation numbers \cite{s3}
\beq E[\rho({\bf r}),n_i]= T_k[\rho({\bf r}),n_i]+ \frac{1}{2}\int
V_H({\bf r})\rho({\bf r})d({\bf r})+ \int V_{ext}({\bf
r})\rho({\bf r})d({\bf r})+ E_{xc}[\rho({\bf r}),n_i].\eeq Here
$T_k$ is the functional of the kinetic energy of non-interacting
Kohn-Sham particles.  Upon substituting eq. (16) into (15) and
using eqs. (3,8), one gets, that the single particle energies
$\epsilon_i$ can be presented by the following expression, \beq
\epsilon_i=\varepsilon_i-<\phi_i|v_{xc}|\phi_i>
-\frac{1}{2}\frac{\delta}{\delta n_i} \int \left[\frac{\chi({\bf
r}_1,{\bf r}_2,iw)+ 2\pi\rho({\bf r}_1)\delta(w)\delta({\bf
r}_1-{\bf r}_2)} {|{\bf r}_1-{\bf r}_2|}\right] \frac{dwdg'd{\bf
r}_1d{\bf r}_2}{2\pi}. \eeq The first and second terms on the
right in eq. (17) is linked with the derivative of functional
$T_k$ with respect to occupation numbers $n_i$. To calculate the
derivative we consider an auxiliary system of non-interacting
particles moving in a field $U({\bf r})$. Then its ground state
energy $E_0$ is given by the following equation \beq E_0=T_k+\int
U({\bf r})\rho({\bf r})d{\bf r}.\eeq Varying  $E_0$ with respect
to the occupation numbers one gets the desirable result, \beq
\varepsilon_i=\frac{\delta E_0}{\delta n_i} =\frac{\delta
T_k}{\delta n_i}+<\phi_i|U|\phi_i>, \eeq provided
$U=V_H+V_{ext}+v_{xc}$. The third term is related to the
contribution coming from $E_{xc}$ defined by eq. (3). The
derivative $\delta\chi/\delta n_i$, as it follows from eq. (4), is
given by, \beq \frac{\delta\chi}{\delta n_i}=
\frac{\delta\chi_0}{\delta n_i}+ \frac{\delta\chi_0}{\delta n_i}R
\chi+\chi_0\frac{\delta R}{\delta n_i} \chi+\chi_0
R\frac{\delta\chi} {\delta n_i}.\eeq In eq. (20) for the sake of
brevity we omit the spatial integrations. The variational
derivative $\delta\chi_0/\delta n_i$ is calculated directly from
(9) and has the simple functional form, \beq
\frac{\delta\chi^0({\bf r}_1,{\bf r}_2,\omega)} {\delta n_i}
=\left[G_0({\bf r}_1,{\bf r}_2,\omega +\varepsilon_i) +G_0({\bf
r}_1,{\bf r}_2,-\omega +\varepsilon_i)\right] \phi^{*}_i({\bf
r}_1) \phi_i({\bf r}_2),\eeq with $G_0({\bf r}_1,{\bf
r}_2,\omega)$ being the Green function of $N$ non-interacting
electrons moving in the single particle potential
$V_H+v_{xc}+V_{ext}$. We do believe, and some evidences can be
found in \cite{ksk}, that the contribution coming from $\delta
R/\delta n_i$ is small. Nonetheless even having omitted this term,
we have still to deal with rather complex eq. (17). Therefore, it
is of interest to illustrate the general consideration of the
one-electron energies with a simple and important example, when
only the exchange part $E_{x}$ of the total exchange-correlation
functional is selected to be treated rigorously, using an
approximation for $E_{c}=E_{xc}-E_{x}$.  Taking into account
calculations of $\varepsilon_c$ given in Table 1, and the local
density approximation, one gets, \beq E_c[\rho]=\int\rho({\bf
r})\varepsilon_c(({\bf r}))d{\bf r}.\eeq For the functional
$E_{x}$ we have an exact expression \cite{as,s2}, \beq
E_{x}[\rho]= -\frac{1}{2}\int\left[ \chi_0({\bf r}_1,{\bf r}_2,iw)
+2\pi\rho({\bf r}_1)\delta(w) \delta({\bf r}_1-{\bf r}_2)\right]
\frac{1}{|{\bf r}_1-{\bf r}_2|} \frac{dw}{2\pi}d{\bf r}_1d{\bf
r}_2. \eeq As it is seen from eq. (6), single particle potential
$v_{xc}$ takes the form, \beq v_{xc}({\bf r})=v_{x}({\bf
r})+v_{c}({\bf r}), \eeq with the potentials $v_{x}({\bf r})$ and
$v_{c}({\bf r})$ given by the following expressions \beq
v_{x}({\bf r})=\frac {\delta E_x}{\delta\rho({\bf r})};\,\,
v_{c}({\bf r})=\frac {\delta E_c}{\delta\rho({\bf r})}.\eeq The
potential $v_x({\bf r})$ can be calculated exactly
\cite{as,s2,gl}, while there are quite suitable approximations to
the potential $v_{c}({\bf r})$ (see e.g. \cite{gk}). Presenting
single particle potential $v_{xc}({\bf r})$ in such a way as (25),
we simplify the calculations a lot, preserving at the same time
the asymptotic condition (7), namely $v_{xc}(r\to\infty)\to-1/r$.
This condition is of crucial importance for calculations of
$\varepsilon_i$ related to the Kohn-Sham vacant states (see eq.
(8)). In the same way as eq. (17) was derived, one gets, \beq
\epsilon_i=\varepsilon_i-<\phi_i|v_{x}|\phi_i>
-\frac{1}{2}\frac{\delta}{\delta n_i} \int \left[\frac{\chi_0({\bf
r}_1,{\bf r}_2,iw)+ 2\pi\rho({\bf r}_1)\delta(w) \delta({\bf
r}_1-{\bf r}_2)} {|{\bf r}_1-{\bf r}_2|}\right] \frac{dw\,d{\bf
r}_1d{\bf r}_2}{2\pi}. \eeq After straightforward calculations one
arrives at rather simple result for the single particle spectra,
that are to be compared with experimental results, \beq
\epsilon_i=\varepsilon_i-<\phi_i|v_{x}|\phi_i> -\sum_kn_k\int
\left[\frac {\phi^*_i({\bf r}_1) \phi_i({\bf r}_2) \phi^*_k({\bf
r}_2)\phi_k({\bf r}_1)} {|{\bf r}_1-{\bf r}_2|}\right] d{\bf
r}_1d{\bf r}_2. \eeq single particle levels $\epsilon_i$, given by
eq. (27), resemble the ones that are obtained within the
Hartree-Fock (HF) approximation. If the wave functions $\phi_i$
would be solutions of the HF equations and the correlation
potential $v_{c}({\bf r})$ would be omitted, the energies
$\epsilon_i$ would exactly coincide with their eigenvalues
$\varepsilon_i$. But this is not the case, since $\phi_i$ are
solutions of eq. (8), and the energies $\epsilon_i$ do not
coincide with either HF eigenvalues or with the ones of eq. (8).

In summary, we have shown that it is possible to calculate the
single particle excitations within the framework of DFT. The
developed equations allow for calculations of the
single particle excitation spectra of any multielectron system such
as atoms, molecules and clusters.  We anticipate as well that these
equations when applied to solids will produce a quite reasonable
results for the single particle spectra and energy gap at various
high-symmetry points in the Brillouin zone.  We have also related
the eigenvalues of the single particle Kohn-Sham equations to the
real single particle spectrum. In the most straightforward case, when
the exchange functional is treated rigorously while the correlation
functional is taken in the local density approximation, the coupling
equations are very simple. The single particle spectra do not
coincide either with the eigenvalues of the Kohn-Sham equations or
with the ones of the Hartree-Fock equations, even provided the
contribution coming from the correlation functional is omitted.

This research was funded in part by INTAS under Grant No.
INTAS-OPEN 97-603.\\

\end{document}